\documentclass [12pt]{article}
\usepackage {graphicx}

\begin{document}
\begin{center}
\textbf{Multifractal Properties of the Ukraine Stock Market}
\end{center}

\begin{center}
A.Ganchuk, V.Derbentsev, V. Soloviev
\end{center}

\begin{center}
Economical Cybernetics Department, Kryviy Rih Economic Institute
Kyiv National Economic University by Vadim Getman, Kryviy Rih,
Ukraine

fax, phone: +380~564 901512\\

v{\_}n{\_}soloviev@kneu.dp.ua
\end{center}

Abstract

Recently the statistical characterizations of financial markets
based on physics concepts and methods attract considerable
attentions. We used two possible procedures of analyzing
multifractal properties of a time series. The first one uses the
continuous wavelet transform and extracts scaling exponents from
the wavelet transform amplitudes over all scales. The second
method is the multifractal version of the detrended fluctuation
analysis method (MF-DFA). The multifractality of a time series we
analysed by means of the difference of values singularity stregth
$\alpha _{\max } $ and $\alpha _{\min } $ as a suitable way to
characterise multifractality. Singularity spectrum calculated from
daily returns using a sliding 1000 day time window in discrete
steps of 1\ldots 10 days. We discovered that changes in the
multifractal spectrum display distinctive pattern around
significant ``drawdowns''. Finally, we discuss applications to the
construction of crushes precursors at the financial markets.

Key words: Multifractal, stock market, singularity spectrum

PACS: 89.20.-a, 89.65.Gh, 89.75.-k

\begin{center}
1. Introduction
\end{center}

Multivariate time series are detected and recorded both in experiments and
in the monitoring of a wide number physical, biological and economic
systems. A first instrument in the investigation of multivariate time series
is the correlation matrix. The study of the properties of the correlation
matrix has a direct relevance in the investigation of mesoscopic physical
systems, high energy physics, investigation of microarray data in biological
systems and econophysics [1].

Quantifying correlations between different stocks is a topic of interest not
only for scientific reasons of understanding the economy as a complex
dynamical system, but also for practical reasons such as asset allocation
and portfolio risk estimation [2--5]. Unlike most physical systems, where
one relates correlations between subunits to basic interactions, the
underlying ``interactions'' for the stock market problem are not known.

Recent empirical and theoretical analysis have shown that this information
can be detected by using a variety of methods. In this paper we used some of
these methods based on Random Matrix Theory (RMT) [6], correlation based
clustering, topological properties of correlation based graph and
multifractal analyses [7] .

In this paper the different aspects of multiscale properties Ukraine stock
market are discussed. The so-called financial stylized facts comprising,
among others, the non-negligible fat tails of log-return distributions,
volatility clustering and its long-time correlations, anomalous diffusion
etc. counter that the financial dynamics is more complex than it is commonly
assumed can also be inferred from a number of recently-published papers
discovering and exploring the multifractal characteristics of data from the
stock markets.

The concept of multifractality was developed in order to describe the
scaling properties of singular measures and functions which exhibit the
presence of various distinct scaling exponents in their different parts.
Soon the related formalism was successfully applied to characterize
empirical data in many distant fields like turbulence, earth science,
genetics, physiology and, as already mentioned, in finance [1].

In the present paper we analyze data from the Ukraine stock market focusing
on their fractal properties. We apply both on the multifractal detrended
fluctuation analysis and on the which are a well-established methods of
detecting scaling behaviour of signals.

\begin{center}
2. Methods and data
\end{center}

Our analysis was performed on the time series of tick-by-tick recordings for
daily returns of all stocks extracted from database time series of prices
the First Stock Trade System (FSTS) index (www.kinto.com) for the ten-year
period 1997-2006. For comparison similar analyses was conducted for of
Russian stock market (Russian Trade System (RTS) -- www.rts.com). The daily
indices of the FSTS and RTS is the largest markets in Ukraine and Russia
consisting of stocks from various sectors. Indices are basically an average
of actively traded stocks, which are weighted according to their market
value.

There are two possible procedures of analyzing multifractal properties of a
time series. The first one uses the continuous wavelet transform and
extracts scaling exponents from the wavelet transform amplitudes over all
scales. This wavelet transform modulus maxima (WTMM) method [8] has been
proposed as a mean field generalized multifractal formalism for fractal
signals. We first obtain the wavelet coefficient at time t0 from the
continuous wavelet transform defined as:

\[
W_a (t_0 ) \equiv a^{ - 1}\sum\limits_{t = 1}^N {p(t)} \psi ((t - t_0 ) /
a)
\]

\noindent
where $p(t)$ is the analyzed time series, $\psi $ is the analyzing wavelet
function, $a$ is the wavelet scale (i.e., time scale of the analysis), and
$N$ is the number of data points in the time series. For $\psi $ we use the
third derivative of the Gaussian, thus filtering out up to second order
polynomial trends in the data. We then choose the modulus of the wavelet
coefficients at each point t in the time series for a fixed wavelet scale
$a$.

Next, we estimate the partition function

\[
Z_q (a) \equiv \sum\limits_i {\left| {W_a (t)} \right|^q}
\]

\noindent
where the sum is only over the maxima values of $\left| {W_a (t)} \right|$,
and the powers $q$ take on real values. By not summing over the entire set
of wavelet transform coefficients along the time series at a given scale $a$
but only over the wavelet transform modulus maxima, we focus on the fractal
structure of the temporal organization of the singularities in the signal.
We repeat the procedure for different values of the wavelet scale $a$ to
estimate the scaling behavior

\[
Z_q (a) \propto a^{\tau (q)}.
\]

In analogy with what occurs in scale-free physical systems, in which
phenomena controlled by the same mechanism over multiple time scales are
characterized by scale-independent measures, we assume that the
scale-independent measures, $\tau (q)$, depend only on the underlying
mechanism controlling the system.

Altrnatively procedure is the multifractal version of the detrended
fluctuation analysis method (MF-DFA) [9]. Given the time series of price
values $p_s (t_s (i)),i = 1,...,N_s $ of a stock $s$s recorded at the
discrete transaction moments $t_s (i)$, one may consider logarithmic price
increments (or returns) $g_s (i) = \ln (p_s (i + 1)) - \ln (p_s (i))$. For
the time series of the log-price increments $G_s \equiv \left\{ {g_s (i)}
\right\}$ one needs to estimate the signal profile

\begin{equation}
\label{eq1}
Y(i) = \sum\limits_{k = 1}^i {(g_s (k) - < g_s > ),i = 1,...,N_s }
\end{equation}

\noindent
where <\ldots > denotes the mean of $G_s $. $Y(i)$ is divided into $M_s $
disjoint segments of length $n$ starting from the beginning of $G_s $. For
each segment $\nu ,\nu = 1,...,M_s ,$ the local trend is to be calculated by
least-squares fitting the polynomial $P_\nu ^{(l)} $ of order $l$ to the
data, and then the variance

\begin{equation}
\label{eq2}
F^2(\nu ,n) = \frac{1}{n}\sum\limits_{j = 1}^n {\left\{ {Y\left[ {(\nu - 1)n
+ j} \right] - P_\nu ^{(l)} (j)} \right\}^2} .
\end{equation}

In order to avoid neglecting data points at the end of $G_s $ which do not
fall into any of the segments, the same as above is repeated for $M_s $
segments starting from the end of $G_s $. The polynomial order $l$ can be
equal to 1 (DFA1), 2 (DFA2), etc. The variances (\ref{eq2}) have to be averaged over
all the segments $\nu $ and finally one gets the $q$th order fluctuation
function

\begin{equation}
\label{eq3}
F_q (n) = \left\{ {\frac{1}{2M_s }\sum\limits_{\nu = 1}^{2M_s } {\left[
{F^2(\nu ,n)} \right]^{q / 2}} } \right\}^{1 / q},q \in R.
\end{equation}

In order to determine the dependence of $F_q $ on $n$, the function $F_q
(n)$ has to be calculated for many different segments of lengths $n$.

If the analyzed signal develops fractal properties, the fluctuation function
reveals power-law scaling

\begin{equation}
\label{eq4}
F_q (n) \propto n^{\tau (q)}
\end{equation}

\noindent
for large n. The family of the scaling exponents $\tau (q)$ can be then
obtained by observing the slope of log-log plots of $F_q $ vs. $n$. $\tau
(q)$ can be considered as a generalization of the Hurst exponent $H$ with
the equivalence $H \equiv \tau (\ref{eq2})$. Now the distinction between monofractal
and multifractal signals can be performed: if $\tau (q) = H$ for all $q$,
then the signal under study is monofractal; it is multifractal otherwise. By
the procedure, $\tau (q),q < 0$ describe the scaling properties of small
fluctuations in the time series, while the large ones correspond to $\tau
(q),q > 0$. It also holds that $\tau (q)$ is a decreasing function of $q$.

By knowing the spectrum of the generalized Hurst exponents, one can
calculate the singularity strength $\alpha $ and the singularity spectrum
$f(\alpha )$ using the following relations:

\begin{equation}
\label{eq5}
\alpha = \tau (q) + q{\tau }'(q),
\quad
f(\alpha ) = q\left[ {\alpha - \tau (q)} \right] + 1,
\end{equation}

\noindent
where ${\tau }'(q)$ stands for the derivative of $\tau (q)$ with respect to
$q$.

\begin{center}
3. Results
\end{center}

Figure 1 shows the First Stock Trade System index and Russia Trade
System index, from 1997 to 2006.
\begin{center}
\begin{figure}[htbp]
\centerline{\includegraphics[width=6.62in,height=5.17in]{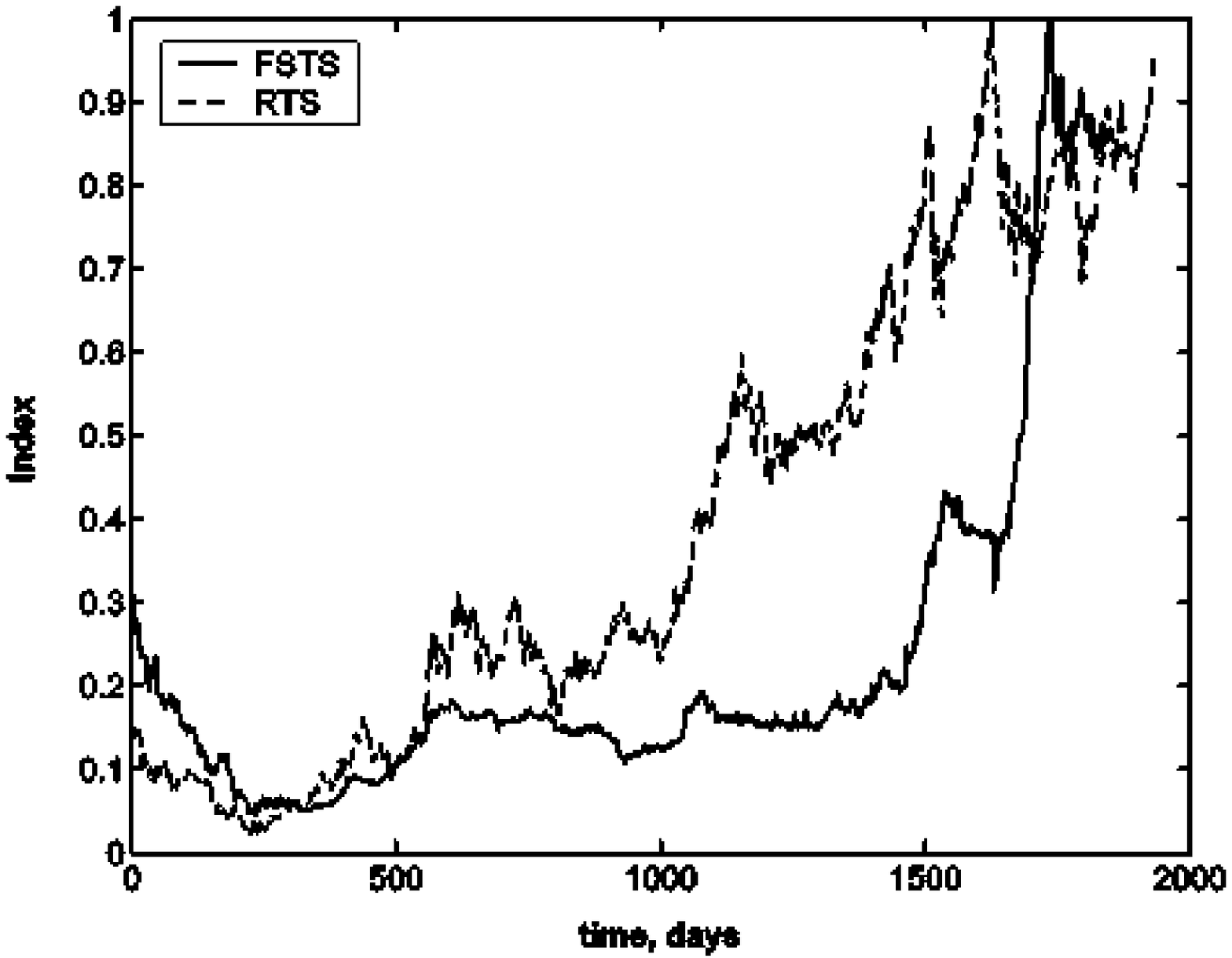}}
\label{fig1}Fig. 1. FSTS and RTS indexes plotted for all the days
reported in the period 1997 to 2006
\end{figure}
\end{center}
Our calculations indicate that the time series of price increments
for all companies can be of the multifractal nature (see fig.2).
Consistently with the log-log plots, the highest nonlinearity of
the spectrum and the strongest multifractality are attributes of
RTS (Russia), and the smallest nonlinearity and the weakest
multifractal character correspond to FSTS (Ukraine).
\begin{table}[htbp]
\begin{tabular}
{p{100pt} p{0pt}p{234pt}p{63pt}p{63pt}}
\multicolumn{2}{p{100pt}}{\centerline{\includegraphics[width=3.0in,height=2.5in]{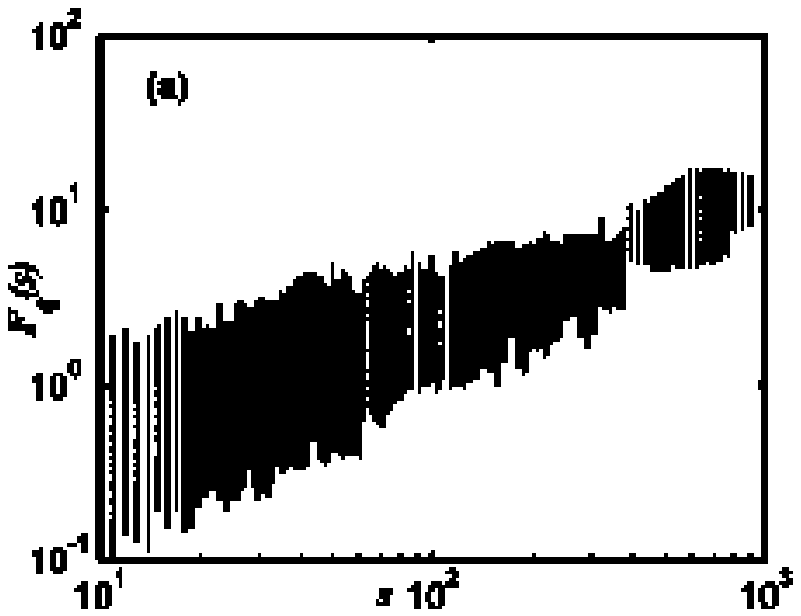}}
\par } &
\multicolumn{3}{p{450pt}}{\centerline{\includegraphics[width=3.0in,height=2.5in]{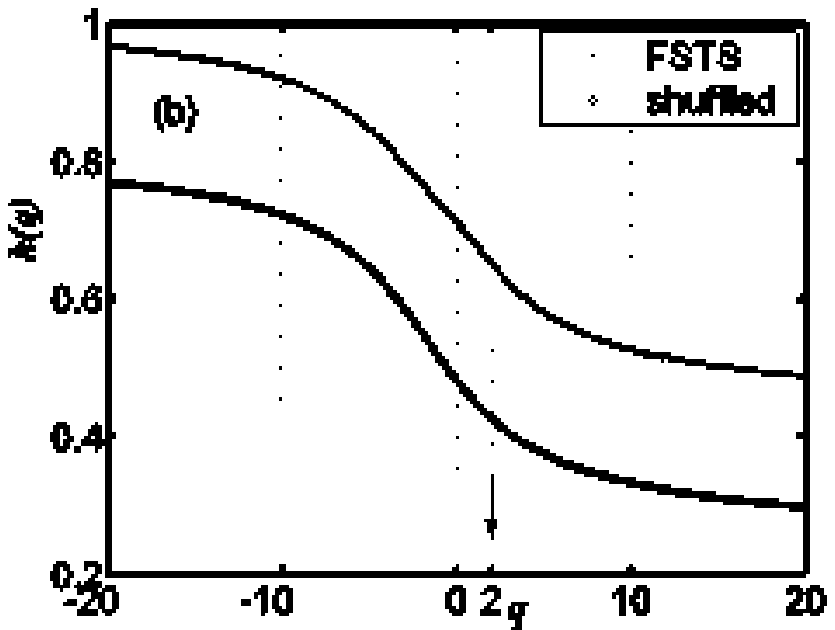}} \par }  \\
\centerline{\includegraphics[width=3.0in,height=2.5in]{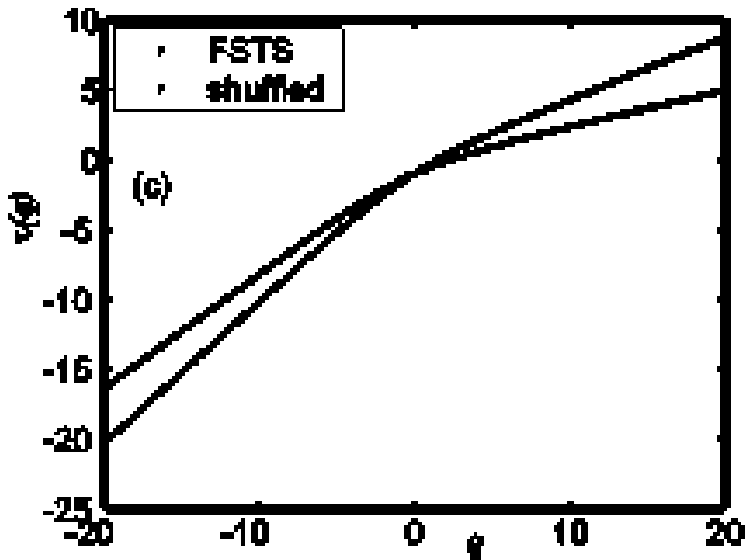}}
\par &
\multicolumn{2}{p{450pt}}{\centerline{\includegraphics[width=3.0in,height=2.5in]{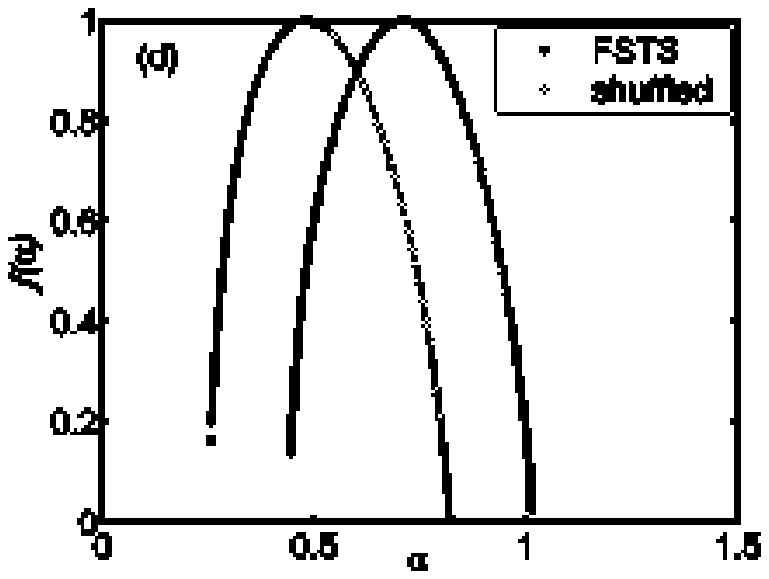}}
\par } &
\multicolumn{2}{ p{450pt} }{}  \\
%\cline{1-3}
\end{tabular}
\label{tab1}Fig. 2. (a) Log-log plots of the $q$-th order
fluctuation $F_q $ for time series of price increments as a
function of segment size n for different values of q between -20
(bottom line) and 20 (top line). (b)Scaling regions allow one to
estimate $h(q)$ according to Eq. 4. (c) Multifractal spectra for
price increments; a nonlinear behaviour of $\tau (q)$ can be
considered a manifestation of multiscaling. (d) singularity
spectra $f(\alpha )$ according to Eq. 5. Open circles corresponds
to shuffled data
\end{table}
The multifractal nature of the data can also be expressed in a
different manner, i.e. by plotting the singularity spectra
$f(\alpha )$(Eq. 5). It is a more plausible method because here
one can easily assess the variety of scaling behaviour in the
data. The evolution of $f(\alpha )$ is analyzed by using a moving
window of length 1000 data point shifted by 1 point. Such a window
ensures that we obtain statistically reliable results.\\
The maxima of $f(\alpha )$are typically placed in a close vicinity
of $\alpha $ = 0.5 indicating no significant autocorrelations
exist. The multifractal character of price fluctuations can
originate from the existence of the long-range correlations in the
price increments (via volatility) as well as from their
non-Gaussian distributions.\\
The richest multifractality (the widest $f(\alpha )$curve $\Delta
\alpha = \alpha _{\max } - \alpha _{\min } )$ is visible for RTS,
the poorest one for FSTS (fig. 3).
\begin{center}
\begin{figure}[htbp]
\centerline{\includegraphics[width=6.63in,height=5.11in]{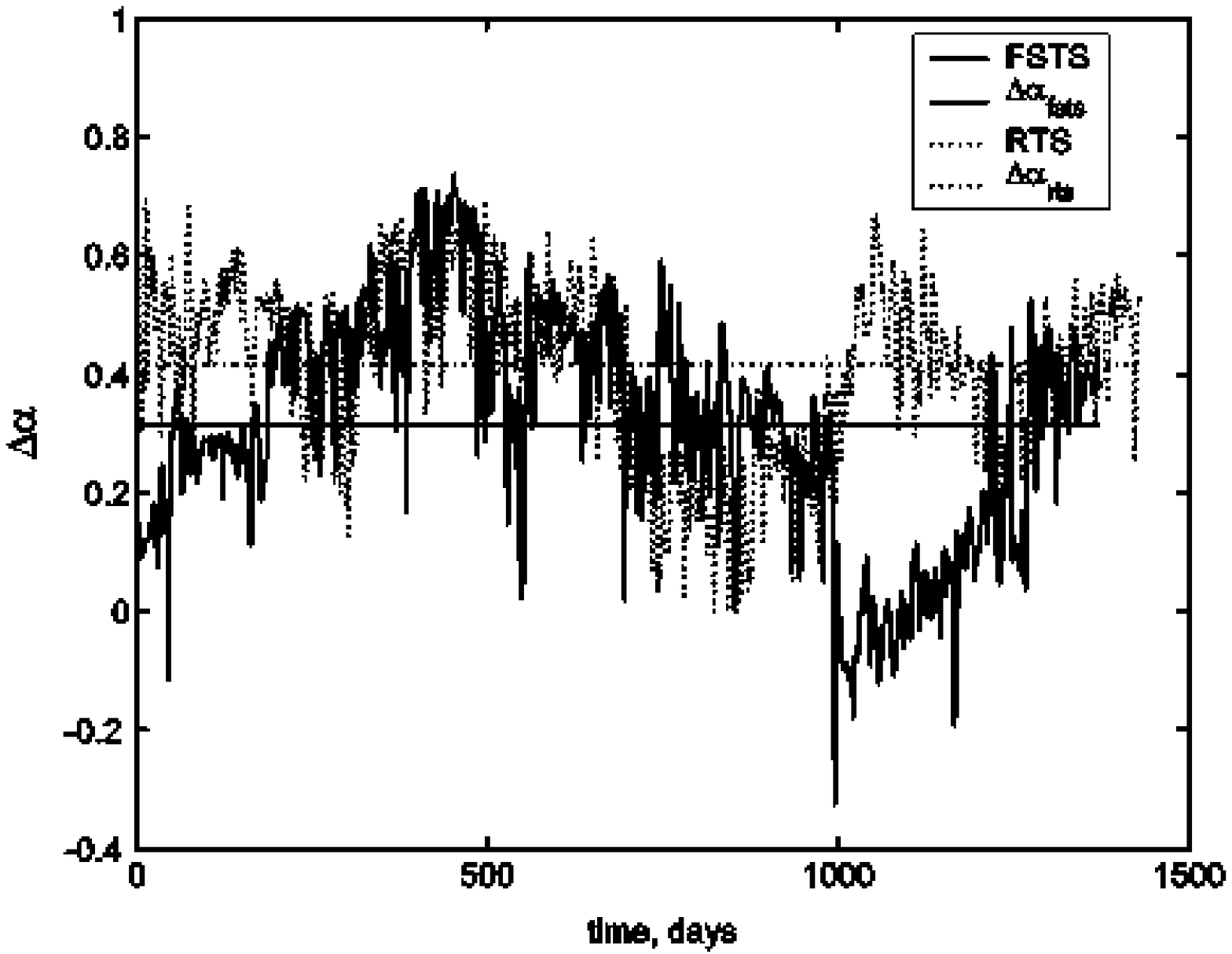}}
\label{fig3}Fig. 3. Comparison of the widths of the $f(\alpha )$
spectra Russia and Ukraine stock markets
\end{figure}
\end{center}

\begin{center}
4. Conclusions
\end{center}

We study the multifractal properties of Ukraine Stock Market. We
show that the signals for the price increments exhibit the
characteristics that can be interpreted in terms of
multifractality. Its degree expressed by the widths $\Delta \alpha
= \alpha _{\max } - \alpha _{\min } )$ of the singularity spectra
$f(\alpha )$different for the Russian and Ukrainian stock markets.
Greater value for the Russian market related to more effective
functioning. In this case the width of singularity spectrum can
serve as the measure of efficiency of functioning of the complex
system.

References

[1] S. Boccaletti$,$ V. Latora, Y. Moreno, M. Chavez , D.-U. Hwang, Physics
Reports 424 (2006) 175 -- 308

[2] B.B. Mandelbrot and J.W. van Ness, SIAM Review 10 (1968) 422-437

[3] V. Plerou, P. Gopikrishnan, L.A.N. Amaral, M. Meyer and H.E. Stanley,

Phys. Rev. E 60 (1999) 6519-6529

[4] P. Gopikrishnan, V. Plerou, L.A.N. Amaral, M. Meyer and H.E. Stanley,

Phys. Rev. E 60 (1999) 5305-5316

[5] V. Plerou, P. Gopikrishnan, L.A.N. Amaral, X. Gabaix and H.E. Stanley,

Phys. Rev. E 62 (2000) R3023-R3026

[6] V. Plerou, P. Gopikrishnan, B.Rosenow, L. A. N. Amaral, T.Guhr, H. E.
Stanley Phys.Rev E, 65 (2002) 066126

[7] S. Drozdz, J. Kwapien, F. Gruemmer, F. Ruf and J. Speth, Acta Phys. Pol.
B 34 (2003) 4293-4305

[8] J. F.Muzy, E. Bacry, and A. Arneodo, Phys. Rev. Lett. \textbf{67
}(1991), 3515

[9] J.W. Kantelhardt, S.A. Zschiegner, E. Koscielny-Bunde, A. Bunde, Sh.
Havlin and H.E. Stanley, Physica A 316 (2002) 87-114

\end{document}